\documentclass[preprint]{vgtc}               

\ifpdf
  \pdfoutput=1\relax                   
  \usepackage{graphicx}                
  \DeclareGraphicsExtensions{.pdf,.png,.jpg,.jpeg} 
\else
  \usepackage{graphicx}                
  \DeclareGraphicsExtensions{.eps}     
\fi%

\graphicspath{{figures/}{pictures/}{images/}{./}} 

\usepackage{microtype}                 
\PassOptionsToPackage{warn}{textcomp}  
\usepackage{textcomp}                  
\usepackage{mathptmx}                  
\usepackage{times}                     
\usepackage{cite}                      
\usepackage{tabu}                      
\usepackage{booktabs}                  

\usepackage{lipsum}                    
\usepackage{mwe}                       
\usepackage{gensymb}
\usepackage{color}
\usepackage{xcolor,colortbl} 
\usepackage{balance}
\usepackage{comment}

\usepackage{mathptmx}                  
\usepackage{soul}
\usepackage{todonotes}
\usepackage{lipsum}

\usepackage{tabularx}
 \newcolumntype{C}{>{\centering\arraybackslash}X}

\onlineid{2003}

\vgtccategory{Research}

\vgtcinsertpkg

\title{Remote Monitoring and Teleoperation of Autonomous Vehicles \\ ---Is Virtual Reality an Option?}

\author{
 Snehanjali Kalamkar$^{1}$\thanks{e-mail: snehanjali.kalamkar@hs-coburg.de}
 \and Verena Biener$^{1}$\thanks{e-mail: verena.biener@hs-coburg.de}
\and Fabian Beck$^{2}$\thanks{e-mail: fabian.beck@uni-bamberg.de}
\and Jens Grubert$^{1}$\thanks{e-mail: jens.grubert@hs-coburg.de}
}
\affiliation{\scriptsize $^{1}$Coburg University of Applied Sciences and Arts, Germany\\  $^{2}$University of Bamberg, Germany}

\teaser{
  \centering
  \vspace*{-0.4cm}
  \includegraphics[width=\linewidth]{figures/Teaser_Study1_Unblurred.png}
  \vspace*{-0.7cm}
  \caption{Conditions of Study 1, from left to right: The physical \textsc{Baseline} interface used in industry in the \textsc{Monitoring} scenario. The replicated \textsc{VR} interface in the \textsc{Monitoring} scenario. The physical \textsc{Baseline} interface in the \textsc{Teleoperation} scenario. The replicated \textsc{VR} interface in the \textsc{Teleoperation} scenario.}
  \label{fig:teaser}
}

\abstract{

While the promise of autonomous vehicles has led to significant scientific and industrial progress, fully automated, SAE level 5 conform cars will likely not see mass adoption anytime soon. Instead, in many applications, human supervision, such as remote monitoring and teleoperation, will be required for the foreseeable future. While Virtual Reality (VR) has been proposed as one potential interface for teleoperation, its benefits and drawbacks over physical monitoring and teleoperation solutions have not been thoroughly investigated. To this end, we contribute three user studies, comparing and quantifying the performance of and subjective feedback for a VR-based system with an existing monitoring and teleoperation system, which is in industrial use today. Through these three user studies, we contribute to a better understanding of future virtual monitoring and teleoperation solutions for autonomous vehicles. The results of our first user study (n=16) indicate that a VR interface replicating the physical interface does not outperform the physical interface. It also quantifies the negative effects that combined monitoring and teleoperating tasks have on users irrespective of the interface being used. The results of the second user study (n=24) indicate that the perceptual and ergonomic issues caused by VR outweigh its benefits, like better concentration through isolation.
The third follow-up user study (n=24) specifically targeted the perceptual and ergonomic issues of VR; the subjective feedback of this study indicates that newer-generation VR headsets have the potential to catch up with the current physical displays.
} 


\begin{document}


\firstsection{Introduction}

\maketitle


Despite massive investments in autonomous driving with certain technological progress, fully autonomous driving (SAE level 5 \cite{sae2018taxonomy}) does not seem achievable in the foreseeable future. Hence, until the introduction of robust level-5 driving, autonomous vehicles (AVs) will require human supervision and direct intervention in certain situations (SAE level 4), as required by law in some countries\footnote{e.g., in Germany through the “Federal Act Amending the Road Traffic Act and the Compulsory Insurance Act" and the “Ordinance on the Approval and Operation of Motor Vehicles with Autonomous Driving Functions in Specified Operating Areas -- Autonomous Vehicles Approval and Operation Ordinance (AFGBV)"}. However, one-on-one supervision---no matter if in the car or remotely---is costly and might hinder the broad application of the technology. Hence, it is relevant to study scenarios where one supervisor can monitor multiple vehicles at once and, if needed, takes over one vehicle to teleoperate it. In fact, this is one mode of operation in industrial use today in Germany. Still, monitoring of the remaining vehicles needs to be continued in these situations. Challenges arise from providing interfaces for a control station that support---although going along with quite different requirements---both monitoring and teleoperation tasks alike. On the one hand, remote monitoring involves tasks like regularly inspecting fuel status, the current location of the AVs, network connection status, etc. On the other hand, teleoperation requires having a cockpit setup available remotely.

Virtual Reality (VR) has already been proposed for the teleoperation of various sorts of vehicles and machines, including robots~\cite{whitney2018ros}, vessels~\cite{ji2020design}, mine-site vehicles~\cite{bednarz2011applications}, and road vehicles~\cite{georg2019adaptable, hosseini2016enhancing, 10.1109/itsc.2018.8569408, shen2016teleoperation}. Likewise, solutions exist that use VR in monitoring scenarios, for instance, in the maritime industry~\cite{tsigkounis2021monitoring}. However, studies of VR interfaces for joint monitoring and teleoperation of AVs or monitoring of AV fleets are yet underexplored.

Hence, within this paper, we quantified the potential benefits and limitations of a physical control interface in industrial use today, which is designed for remote monitoring and teleoperation of AVs, to an interface replicated one-to-one in VR. This alone would provide evidence of whether a replicated VR setup would be a cost-efficient and portable alternative to the physical control station.

Specifically, the paper comprises three user studies implemented as controlled experiments shown in \autoref{fig:teaser}, \autoref{fig:Study2-Bl-vs-Op}, and \autoref{fig:Study3}. An initial user study (n=16) compares an existing physical interface and a replicated VR setup.
While the joint monitoring and teleoperation setup has the potential benefit of being able to support two task types, our initial study quantifies the substantial performance costs in joint monitoring and teleoperation (ca. a quarter more missed alerts and a quarter longer reaction times). 

Hence, in the second study (n=24), we investigated a screen layout that was optimized for a single task (monitoring). 
Specifically, we compared the existing baseline layout (used for monitoring and teleoperation) with a layout optimized for monitoring only, displayed on a physical screen and in VR.
The subjective feedback from this study, majorly was a lower perceived performance due to the limitations of the VR headset like peripheral blur, and the weight of the headset. Hence, we conducted a third study (n=24) to quantify the performance for the same task with an ergonomically and technologically improved VR headset.

In summary, our contributions and main results from the three studies are the following: 
1) For monitoring tasks without teleoperation, we observe comparable reaction times in the baseline (the interface used in industry) and the replicated VR interface, but 7.6\% more missed alerts in the replicated VR interface. 
2) Compared to monitoring without teleoperation, we quantify that parallel monitoring and teleoperation lead to worse monitoring performance in terms of 27.4\% more missed alerts in the baseline and 26.1\% in the replicated VR interface. 
3) Upon optimizing the interface layout specifically for the monitoring task, we observe 11.6\% shorter reaction times in the optimized physical layout as compared to the non-optimized physical layout (baseline). 
However, this observation does not translate to the optimized VR interface. 
4) We observe a 16.5\% higher task load in the replicated VR interface as compared to the baseline. On optimization, we observe approximately 15.1\% lower task load in the optimized physical interface as compared to the baseline. Again, this observation does not translate to the optimized VR interface. 
5) Finally, we observed an equivalence of reaction times between the physical display and the Meta Quest Pro, unlike the second user study. Also, 50\% of the participants preferred using the Meta Quest Pro over the physical display indicating the perceived benefits of an improved VR headset.

\vspace{-0.3cm}
\section{Related Work}

With the dawn of (partially) autonomous vehicles and cost-efficient VR headsets, VR has been investigated as a tool for automated driving research such as passenger-focused experiences, e.g., for productivity and leisure activities \cite{grubert2018office, mcgill2020challenges, li2021rear}, user interface and experience design for automated driving \cite{morra2019building, paredes2018driving, hensch2020effects, jansen2023autovis} or for vulnerable road users \cite{ye2020risks, colley2020effect, colley2022effects}. For a recent overview, we refer to a survey by Riegler et al. \cite{riegler2021systematic}. 

A variety of applications requires the use of control centers for remote monitoring and teleoperation. Bergroth et al. \cite{bergroth2018use} explored the use of VR control centers for individual monitoring and control tasks in the context of nuclear power plants. Fabris et al. \cite{fabris2021immersive} investigated the impact of immersive telepresence on the operation of unmanned vehicles. Kalinov et al. \cite{kalinov2021warevr} proposed a VR application for natural human interaction with an autonomous robotic system for stocktaking, monitoring, and teleoperation. The results of their user study indicate a better performance of the VR interface as compared to a first-person view mode shown on a desktop display. Tsigkounis et al. \cite{tsigkounis2021monitoring} demonstrated an immersive VR dashboard for a marine vessel environment, their user study implied the potential of VR to replace physical large-display dashboards. This concept can be extended to the control centers for autonomous driving, as these require 1:1 monitoring and might require human assistance in terms of decision-making or teleoperation. 
Our work focuses on exploring if and how to utilize VR for monitoring multiple AVs and supporting their teleoperation.

Teleoperation of AVs is necessary in case of critical scenarios.
Neumeier et al. \cite{neumeier2018way} presented a teleoperation station making use of physical display monitors to show the live camera feed from the AV and other potentially relevant information like speed, direction, etc. They also suggested the use of VR for this purpose. Shen et al. \cite{shen2016teleoperation} demonstrated the possibilities of immersive teleoperation of a physical test vehicle, using commercial off-the-shelf components and low latency teleoperation performance with wireless technologies like IEEE 802.11n WiFi, 3G, and 4G.  Hosseini and Lienkamp \cite{hosseini2016enhancing} introduced a VR interface, illustrating 360$^{\circ}$ surroundings of a remote vehicle using the camera and LiDAR transmitted data. Their evaluations showed a significant improvement in task performance using VR HMI in precise test control scenarios. 
Georg and Diermeyer \cite{georg2019adaptable} proposed an immersive and adaptable interface for the teleoperation of AVs, considering the scalability of components. 
Gafert et al. \cite{gafert2022teleoperationstation} presented TeleOperationStation, an XR prototype for the remote operation of a fleet of AVs. These prior works looked into the teleoperation of AVs in isolation. 
Our work tries to evaluate the feasibility of the handover scenarios where a remotely available human must teleoperate the vehicle for a short duration of time, bringing the vehicle to a stable state, while still monitoring the other vehicles. 

As the possibility of teleoperation of vehicles is still being explored, it is important to focus on user-friendly interfaces for remote monitoring and teleoperation of AVs. Graf et al. \cite{graf2020design} presented a design space to support the development of user interfaces that should possibly improve remote situational awareness. Graf and Hussmann \cite{graf2020user} further presented a comprehensive situational awareness requirements framework and analysis for AVs' teleoperation interfaces. Their analysis resulted in the elicitation of 80 requirements from 12 categories. Kettwich et al. \cite{kettwich2021teleoperation} presented a user-centered human-machine interface in the context of public transport control centers. This was a click-dummy prototype of the potentially designed interface for a 2D display. 
To bring more insights into user interfaces for such AV control centers, our work evaluates the effects of using a virtual replica of an existing AV control center. We also evaluate the usability of VR with an interface optimized specifically for monitoring purposes.

Lischke et al. \cite{lischke2016screen} explored a variety of screen arrangements for multi-monitor setups and classified them based on user preferences. These arrangements could translate to mixed reality environments, too.
The impacts of using virtual monitors displayed in mixed reality have been explored by Pavanatto et al. \cite{pavanatto2021we}. Their findings indicate that virtual monitors can be used for a short period of productivity work. Ens et al. \cite{ens2014personal} explored design space for multi-tasking and switching between multiple windows in head-worn displays. McGill et al. \cite{mcgill2020expanding} developed, evaluated, and exhibited a significantly beneficial technique to minimize the physical effort and discomfort in viewing through HMDs due to the limited field of view. 
Inspired by these prior works, as an initial step, we explore the effects of bringing the multi-monitor physical setup into VR, for a different use case, i.e., remote monitoring and teleoperation of AVs.

\vspace{-0.1cm}
\section{Study 1: VR Replication of a Joint Monitoring and Teleoperation System}

\label{sec:replication-study}
In the initial study, we investigated the effects of using a VR replica of an existing industrial physical monitoring and teleoperation system to quantify the benefits and limitations of the one-to-one VR replica. Further, we quantified the adverse effects on users when they need to divide their attention between two cognitively demanding tasks (teleoperation and monitoring).

To this end, we designed a go/no-go inspired experiment in which participants had to react to relevant stimuli while ignoring distractor stimuli in a monitoring-only and a joint monitoring and teleoperation task.

Study 1 was conducted as a 2$\times$2 within-subjects design with the independent variables \textsc{Interface} and \textsc{Scenario}. 
The two levels of \textsc{Interface} were \textsc{Baseline}, representing the existing physical interface, and \textsc{VR}, representing the physical interface replicated in VR. 
The two levels of \textsc{Scenario} were \textsc{Monitoring}, in which the participants only monitored the AVs, and \textsc{Teleoperation}, in which the participants drove a virtual vehicle while monitoring the AVs.
This led to four conditions (\textsc{Baseline} and \textsc{Monitoring}, \textsc{Baseline} and \textsc{Teleoperation}, \textsc{VR} and \textsc{Monitoring}, \textsc{VR} and \textsc{Teleoperation}).
In each condition, participants had to react to 18 alerts.
As dependent variables, we measured the miss rate (no reaction), reaction time for each individual alert, and error rate. 
In \textsc{Teleoperation}, we also measured the deviation from a prescribed driving path and the distance traveled in both interfaces. 
Additionally, we collected subjective data including usability (system usability scale (SUS) \cite{brooke1996sus}), task load (Raw NASA TLX questionnaire (TLX) \cite{hart1988development}), simulator sickness (simulator sickness questionnaire (SSQ) \cite{kennedy1993simulator}), presence using the IPQ questionnaire \cite{schubert2003sense}. We further conducted semi-structured interviews after each condition and at the end. 
After each condition, we asked the participants “What did you like in this condition? What were the problems?” and after all conditions, “Which condition was the best for you? Why? Could you imagine using VR for monitoring and teleoperation in the future? What would you improve?” 
The order of the four conditions was balanced while blocking for \textsc{Interface}, so they either performed both \textsc{VR} conditions first or both \textsc{Baseline} conditions. 

\vspace{-0.4cm}
\subsection{Apparatus}
\label{section:apparatus-1}
For the user study, we used a physical setup inspired by the existing monitoring and teleoperation system in use today at Valeo, which works with four vehicles simultaneously. 
We built the same physical setup for our user study and replicated it as detailed as possible in our \textsc{VR}-Setup. 
This physical setup \textsc{Baseline} consisted of four monitors mounted on a wall and a curved monitor placed on a table (see \autoref{fig:teaser}). 

The topmost monitor on the wall was a 65-inch TV monitor (JAY-TECH S65U65129M UHD LED TV) referred to as \textsc{top} placed at a height of 234 cm facing downwards at an angle of 15$^{\circ}$. It showed the live locations of all four vehicles on separate maps, as shown in \autoref{fig:teaser}. 
Three identical 32-inch Samsung UHD monitors were placed directly below it, at a height of 146 cm facing straight ahead, with the outer two monitors rotated by 7$^{\circ}$ towards the user. 
The left monitor, referred to as \textsc{left}, displayed the sensor data of all vehicles, like network, connection status, doors, battery, mode of operation, and vehicle type.
The middle monitor, referred to as \textsc{middle}, showed additional sensor data of the selected vehicle like speed, driving mode (auto, teleoperated, manual), and indoor temperature. 
The right monitor, referred to as \textsc{right}, displayed live camera feeds from four vehicles.
The curved monitor was a 49-inch curved monitor (Samsung C49RG94SSU) positioned at a height of 102 cm and displayed the front and rear view from the live camera feeds and the speed of the vehicle to be teleoperated. 

A GTTRACK Racing Simulator Cockpit was placed in front of the monitors. The seat in the simulator cockpit was at a height of 68 cm. The average distance of the participants' eyes from the ground was 120 cm, and that to the curved screen was about 130 cm. These measurements were replicated from the industrial setup. A Thrustmaster T300 steering wheel and corresponding pedals were attached to the racing simulator cockpit. These were used in the \textsc{Teleoperation} conditions to operate the vehicle.

For the \textsc{VR} conditions, we used an HTC Vive Pro 2. In the virtual environment, virtual monitors of the same size and position as in the \textsc{Baseline} interface were presented. 
We used an HTC Vive tracker to track the physical steering wheel and visualize it in VR.

Please note, while the physical monitoring and teleoperation system underlying \textsc{Baseline} condition is normally used to operate actual AVs on public roads, in this study we did not monitor or operate physical vehicles due to the associated real-world risks. Instead, we simulated vehicles using the CARLA simulator for autonomous driving research by Dosovitskiy et al. \cite{Dosovitskiy17}. 

\vspace{-0.2cm}
\subsection{Tasks}
The participants had two kinds of tasks: monitoring and teleoperation (driving). The monitoring task was to react to relevant alerts as quickly as possible and as much as possible while ignoring the non-relevant tasks. The driving task was to drive a virtually simulated vehicle on a pre-defined course while obeying the traffic rules. 

\paragraph{Monitoring Task.}
\label{sec:monitoring-tasks}
In both the interfaces (\textsc{VR} and \textsc{Baseline}), participants 
had to react to visual alerts. Please note, in actual monitoring setups, such alerts could potentially be detected automatically. In our user study, these alerts acted as plausible placeholder items for events that cannot be automated in real-world monitoring systems and were used to explore how well VR could support the perception of visual alerts.
The alerts were shown on one of three screens (\textsc{top}, \textsc{left}, \textsc{right}). For each alert, participants had to decide if it was a relevant alert or not. Participants were required to react only to relevant alerts while ignoring non-relevant ones. 

\begin{figure}[t]
	\centering
	\includegraphics[width=1\columnwidth]{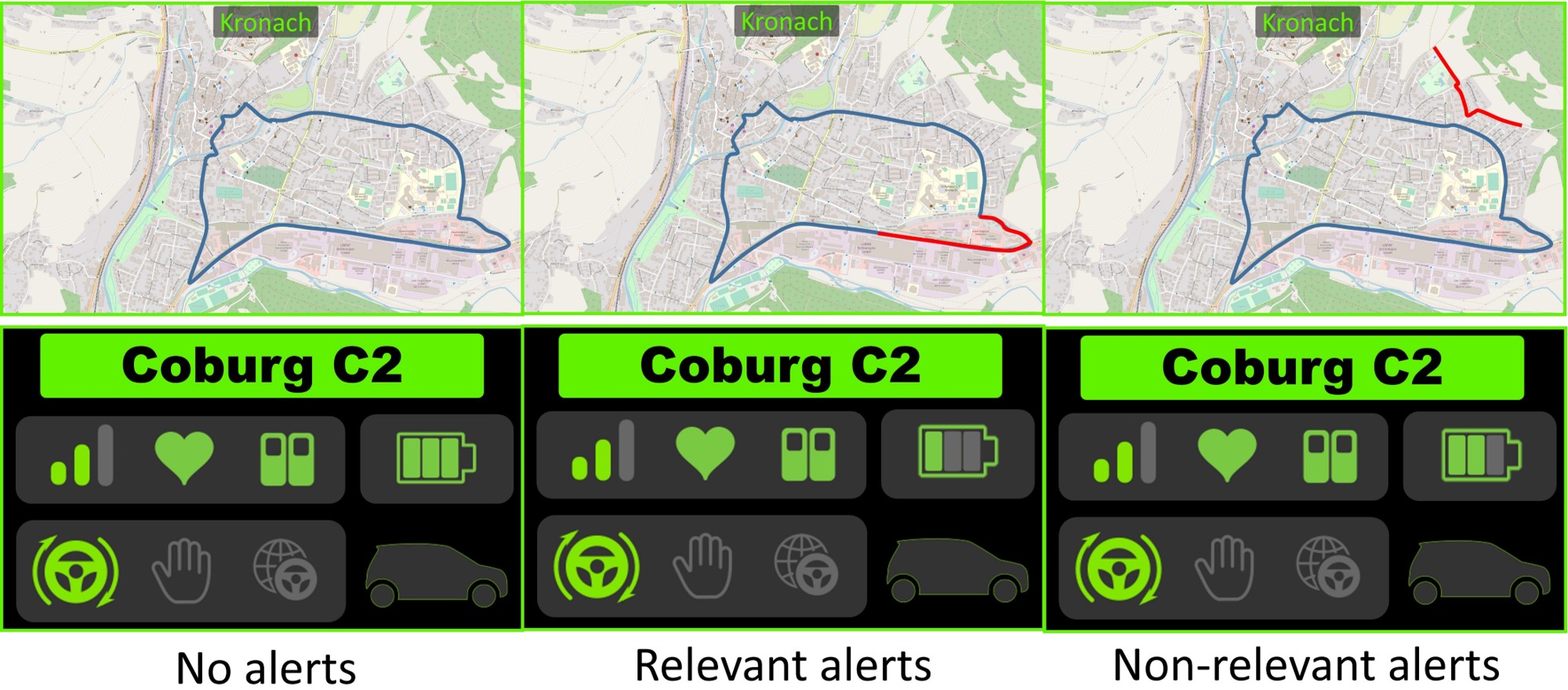}
        \vspace*{-0.6cm}
	\caption{Left column: no alerts. Middle column: relevant map and battery alerts. Right column: non-relevant map and battery alerts.
    }
	\label{fig:map-and-battery-alerts}
 \vspace{-0.6cm}
\end{figure}

On \textsc{top}, 
an alert was displayed by coloring certain street segments red, indicating a high volume of traffic. 
If the indicated traffic volume was on the predefined route of the vehicle (indicated in blue on the map), the participants were asked to treat those alerts as relevant and react by saying the keyword “map".
If the indicated traffic volume was not on the predefined route of the vehicle, they had to ignore it, see \autoref{fig:map-and-battery-alerts}, top row.
On \textsc{left}, 
an alert was indicated by a change in the battery symbol. The battery symbol could either display one, two, or three bars. 
The status change of the battery to one bar was considered relevant, and the participants had to react by saying the keyword “battery", while a change to two bars was not relevant and was to be ignored, see \autoref{fig:map-and-battery-alerts}, bottom row.
On \textsc{right}, it was considered an alert when a camera feed was lost, which was visible as a change in motion (frozen image), or the camera feed updated at a low frame rate. If either of the camera feeds was lost, the participants had to react by saying the keyword “camera", while a drop in frame rate was not relevant and therefore was to be ignored, see Fig. 1 in the supplementary material. 

In a duration of 15 minutes, a total of 36 alerts (18 relevant and 18 non-relevant) were generated in random intervals between 20 to 30 seconds. 
Each alert was visible for 5 seconds (determined empirically) and there were no overlapping alerts. 
Regardless of the type of alert, participants were asked to react to relevant alerts by saying the respective keyword. 
If the participants did not react to a relevant alert within 7 seconds of it being displayed or ignored a relevant alert, it was logged as missed. 
We used a threshold of 7 seconds for a reaction to allow the participants to react even if they saw the alert at the fifth second or if they could not recall the keyword quickly. 
If the participants reacted to a non-relevant alert or reacted with the wrong keyword, it was logged as an error. 

\paragraph{Driving Task.}
In both the interfaces (\textsc{VR} and \textsc{Baseline}), participants performed a driving task in the \textsc{Teleoperation} scenario, where they had to drive a virtually simulated vehicle.
The participants drove in Town01 from the CARLA simulator \cite{Dosovitskiy17} in one \textsc{Interface} and in Town07 in the other. 
The combination of the \textsc{Interface} and the town was balanced among the participants. 
The pre-defined driving path was visualized using spheres (see Fig. 3 in the supplementary material).
We asked the participants to follow this path, drive as fast as possible, and align the position of the vehicle below the spheres as accurately as possible to minimize the deviation from the path. 
We also asked them to avoid the obstacles on the path (see Fig. 4 in the supplementary material), which were placed to increase the difficulty of driving. 
We reminded the participants of obeying all traffic rules like following the speed limit, driving in the appropriate lane, etc., as the virtual simulation was a representative of an actual vehicle with humans on board. 
Since the \textsc{Teleoperation} scenario involved the monitoring tasks, too, we asked the participants to also monitor the alerts, leaving it to their judgment on how to handle the two tasks simultaneously.



\paragraph{Task Simulation.} We implemented an Unreal Engine (UE) 4.27 application server for simulating the alerts and visualizing them in \textsc{VR} and used ReactJS web pages with Node.JS server for visualization in \textsc{Baseline}. 
The implementation details are provided in Section 1.2 of the supplementary material.


\vspace{-0.1cm}
\subsection{Procedure}
\label{sec: participants-1}

We used G*Power \cite{faul2007g} to determine the minimum sample size required for the user study based on the correlation between conditions from our pilot study. 
With a significance level $\alpha = 0.05$, statistical power $1-\beta = 0.8$, and our correlation values $0.75, 0.85, 0.92, 0.88$, the minimum sample size was 13. 
To ensure counterbalancing, we increased it to $n=16$. 
The participants had to be at least 18 years old and have a driver's license to participate in the study. 
Sixteen participants took part in the study (4 females and 12 males), with an average age of 25.19 years ($sd=3.9$), an average height of 178.5 cm ($sd=10.12$), and an average driving experience of 5.35 years ($sd=3.77$). 
Four participants had no experience with VR. Two participants had no experience of car racing games and the others had minimal to moderate experience. The frequency of the participants driving in their daily lives spanned from driving daily to driving once a year. Six participants wore glasses during the user study. 
We asked the participants to rate their confidence level while driving on a scale of 1 (being very insecure) to 10 (being very confident).
This resulted in an average rating of 7.875 ($sd = 1.85$).

Firstly, all participants were informed about the procedure and the content of the study. They signed a consent form and filled out a demographic questionnaire including information about their driving experience. 
Participants were given an introduction to all the interfaces and tasks.
The cockpit was adjusted to the height of each participant by moving the pedals.
They started the first condition with a short training session of up to 5 minutes.
After finishing the first condition which lasted for 15 minutes, they answered the 
subjective questionnaires and semi-structured interview questionnaires.
This procedure was repeated for the other three conditions.
Lastly, they answered the preference questionnaire and the interview questions reflecting on the differences between the conditions. The study lasted for 2 hours per participant.
Each participant was rewarded with a 10 EUR gift card for local businesses as a sign of gratitude.

\vspace{-0.1cm}
\subsection{Results}
\label{sec: results-vr-vs-bl}

\begin{table}[t]
    \centering 
    \tiny
    \caption{
    RM-ANOVA results for Study 1.
    d$f_1$ = d$f_{effect}$ and d$f_2$ = d$f_{error}$. I=\textsc{Interface}. S=\textsc{Scenario}.}
    \setlength{\tabcolsep}{5pt}
        
        \begin{tabularx}{0.47\textwidth}{|C||c|c|c|c|c||c|c|c|c|c|}
            \hline
            &\multicolumn{5}{|c|}{Miss Rate} &\multicolumn{5}{|c|}{Reaction Time} 
            \\
            \cline{2-11}
            & d$f_{1}$ & d$f_{2}$ & F & p &  $\eta^2_p$
            & d$f_{1}$ & d$f_{2}$ & F & p &  $\eta^2_p$
            \\
            \hline
            \textsc{I} & $1$ & $15$ & $7.594$ & \cellcolor{lightgray}$0.015$ & $0.336$
                                 & $1$ & $15$ & $0.047$ & $0.832$ & $0.003$
                                 \\
            \textsc{S} & $1$ & $15$ & $76.73$ & \cellcolor{lightgray}$<0.01$ & $0.836$
                              & $1$ & $15$ & $1.755$ & $0.205$ & $0.105$
                              \\
            \textsc{I} $\times$ \textsc{S} 
                                 & $1$ & $15$ & $0.073$ & $0.791$ & $0.005$
                                 & $1$ & $15$ & $0.014$ & $0.909$ & $0.001$
                                 \\                     
            \hline                    
        \end{tabularx}
        \\ 
        \begin{tabularx}{0.47\textwidth}{|C||c|c|c|c|c||c|c|c|c|c|}
            \hline
            &\multicolumn{5}{|c|}{Error Rate} &\multicolumn{5}{|c|}{System Usability}
            \\
            \cline{2-11}
            & d$f_1$ & d$f_2$ & F & p &  $\eta^2_p$
            & d$f_1$ & d$f_2$ & F & p &  $\eta^2_p$
            \\
            \hline
            \textsc{I}  & $1$ & $15$ & $2.129$ & $0.166$ & $0.124$ 
                                & $1$ & $14$ & $1.135$ & $0.305$ & $0.075$
                                  \\
            \textsc{S} & $1$ & $15$& $0.0165$ & $0.9$ & $0.001$ 
                              & $1$ & $14$ & $14.371$ & \cellcolor{lightgray}$<0.01$ & $0.507$
                              \\
            \textsc{I} $\times$ \textsc{S} 
                                 & $1$ & $15$ & $3.393$ & $0.085$ & $0.184$ 
                                 & $1$ & $14$ & $0.108$ & $0.747$ & $0.008$              
                                 \\                       
            \hline                    
        \end{tabularx}
        \\ 
        \begin{tabularx}{0.47\textwidth}{|C||c|c|c|c|c||c|c|c|c|c|}
            \hline
            &\multicolumn{5}{|c|}{Task Load} &\multicolumn{5}{|c|}{Simulator Sickness} \\
            \cline{2-11}
            & d$f_{1}$ & d$f_{2}$ & F & p &  $\eta^2_p$
            & d$f_{1}$ & d$f_{2}$ & F & p &  $\eta^2_p$
            \\
            \hline
            \textsc{I} & $1$ & $15$ & $7.676$ & \cellcolor{lightgray}$0.014$ & $0.339$
                                 & $1$ & $14$ & $12.566$ & \cellcolor{lightgray}$0.003$ & $0.473$ 
                                 \\
            \textsc{S}  & $1$ & $15$ & $17.82$ & \cellcolor{lightgray}$<0.01$ & $0.543$
                              & $1$ & $14$ & $4.409$ & $0.054$ & $0.234$ 
                              \\
            \textsc{I} $\times$ \textsc{S} & $1$ & $15$ & $2.895$ & $0.109$ & $0.162$
                                 & $1$ & $14$ & $0.006$ & $0.941$ & $<0.01$ 
                                 \\                     
            \hline                    
        \end{tabularx}
        \vspace{-0.4cm}
        \label{tab:resultsTable}
\end{table}

\begin{figure*}[t]
	\centering 
	\includegraphics[width=\linewidth]{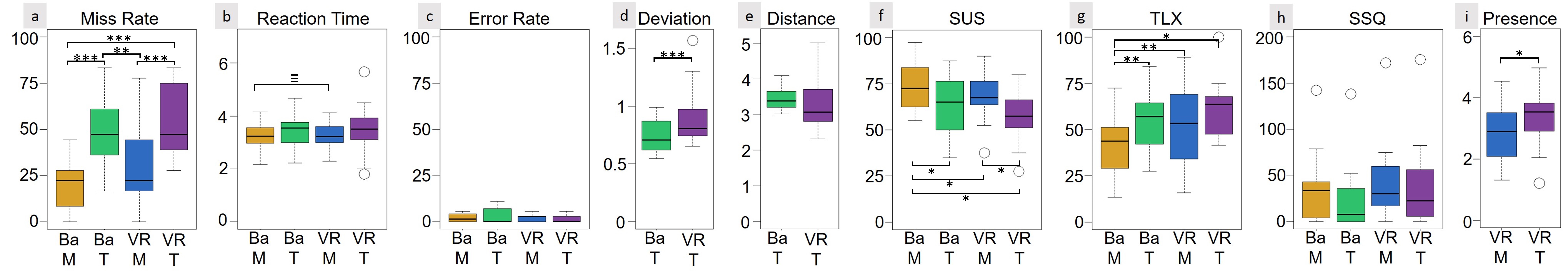}
        \vspace*{-0.7cm}
	\caption{
        Box plots for dependent variables in Study 1: a) Miss Rate. b) Reaction Time (in seconds). c) Error Rate. d) Deviation from Path (in m)  e) Distance Traveled (in km). f) System Usability Scale. g) NASA Task Load Index. h) Simulator Sickness. i) Presence. Ba=\textsc{Baseline}, VR=\textsc{VR}, M=\textsc{Monitoring}, T=\textsc{Teleoperation}. The number of stars indicates the level of significance between the conditions (*** \textless 0.001 **  \textless 0.01 *  \textless 0.05). Equivalence between conditions is indicated by $\equiv$.
        }
	\label{fig:study1-analysis-plots}
 \vspace{-0.6cm}
\end{figure*}

The influence of \textsc{Interface} and \textsc{Scenario} was analyzed using repeated measures analysis of variance (RM-ANOVA). 
We used the Aligned Rank Transform on the non-normal subjective data and then applied RM-ANOVA for analysis. For the non-normal objective data, we used RM-ANOVA, as the sphericity requirements were met and normality violations do not tend to have a major impact on the robustness of the analysis \cite{blanca2023non}.
Bonferroni adjustments were used for multiple comparisons at an initial significance level of $\alpha=0.05$.
We used two one-sided t-tests (TOST) for testing the equivalence between conditions. The equivalence lower and upper bounds were calculated using Cohen's $d_z = \frac{t}{\sqrt{n}}$ with the critical $t$ value for $n=16,\alpha=0.05$ for two-tails (c.f., \cite{lakens2018equivalence}). 
The value of $d_z$ for our analysis was $\pm0.53$, with $-0.53$ as the lower bound and $+0.53$ as the upper bound. 
Due to data logging errors, we lost the SSQ and SUS questionnaire data of one participant. As this missing data does not affect the participant's task data, we did not exclude the participant from the entire statistical analysis. 
The box plots for each dependent variable are shown in \autoref{fig:study1-analysis-plots}.
The results of the statistical tests are presented in \autoref{tab:resultsTable}.
We followed an open coding procedure \cite{corbin1990grounded}, to analyze the interviews. 

\vspace{-0.03cm}
\paragraph{Miss Rate:} The miss rate was the percentage of relevant alerts that a participant missed in a particular condition. 
\textsc{VR} resulted in a significantly higher (7.6\%) miss rate ($m=42.2$, $sd=4.272$) compared to \textsc{Baseline} ($m=34.5$, $sd=3.808$). 
\textsc{Teleoperation} resulted in a significantly higher (26.7\%) miss rate ($m=51.7$, $sd=3.335$) compared to \textsc{Monitoring} ($m=25$, $sd=3.263$). 
There were no interaction effects between \textsc{Interface} and \textsc{Scenario}. 

\vspace{-0.03cm}
\paragraph{Reaction Time:} 
The reaction time (in seconds) was the average of all the reaction times to the relevant alerts that the participant reacted to, in a particular condition.
We did not find a significant effect of \textsc{Interface} while comparing \textsc{VR} ($m=3.39$, $sd=0.737$) to \textsc{Baseline} ($m=3.35$, $sd=0.577$).
We also did not find a significant effect of \textsc{Scenario} while comparing \textsc{Monitoring} ($m=3.27$, $sd=0.478$) to \textsc{Teleoperation} ($m=3.46$, $sd=0.794$). There was no significant interaction effect between \textsc{Interface} and \textsc{Scenario}. 
In the \textsc{Monitoring} scenario, TOST indicated equivalence of reaction time between \textsc{VR} ($m=3.28$, $sd=0.48$) and \textsc{Baseline} ($m=3.26$, $sd=0.492$), with the larger of the two p-values being $p = 0.002$ and corresponding $t(15) = -3.328$. 

\vspace{-0.03cm}
\paragraph{Error Rate:} The error rate was the percentage of all alerts that a participant reacted to wrongly in a condition. We did not find a significant effect of \textsc{Interface} while comparing \textsc{VR} ($m=1.65$, $sd=0.737$) to \textsc{Baseline} ($m=2.6$, $sd=0.577$).
We also did not find a significant effect of \textsc{Scenario} while comparing \textsc{Monitoring} ($m=2.08$, $sd=0.478$) to \textsc{Teleoperation} ($m=2.17$, $sd=0.794$). Neither did we find equivalence between either pair among \textsc{Interface} and \textsc{Scenario} taken together.

\vspace{-0.03cm}
\paragraph{Deviation from Path:}  This metric was the average perpendicular distance (in meters) from the line passing through the two nearest consecutive navigational spheres from the vehicle's position at that instance. \textsc{VR} resulted in a significantly higher ($p<0.001 $, $\eta^2p = -0.974$, 22.5\%) deviation from the path ($m=0.909$, $sd=0.274$) compared to \textsc{Baseline} ($m=0.742$, $sd=0.144$). 

\vspace{-0.03cm}
\paragraph{Distance Traveled:} This metric was the total distance (in kilometers) driven by the participants in the \textsc{Teleoperation} scenario. We did not find a significant effect of \textsc{Interface} while comparing \textsc{VR} ($m=3.286$, $sd=0.781$) to \textsc{Baseline} ($m=3.448$, $sd=0.325$), neither did we find equivalence between \textsc{VR} and \textsc{Baseline} for the distance traveled. 

\vspace{-0.03cm}
\paragraph{System Usability:} We did not find a significant effect of \textsc{Interface} on usability while comparing \textsc{VR} ($m=63.2$, $sd=14.71$) to \textsc{Baseline} ($m=68.2$, $sd=16.03$) and neither did we find equivalence among the two levels of \textsc{Interface} for either \textsc{Scenario}.
For \textsc{Scenario}, however, \textsc{Monitoring} resulted in a significantly higher (19\%) usability ($m=71.3$, $sd=13.562$) compared to \textsc{Teleoperation} ($m=60$, $sd=15.355$). There was no significant interaction effect between \textsc{Monitoring} and \textsc{Teleoperation}. 

\vspace{-0.03cm}
\paragraph{Task Load:} \textsc{VR} resulted in a significantly higher (16.5\%) task load ($m=56.5$, $sd=18.128$) compared to \textsc{Baseline} ($m=48.5$, $sd=17.18$). It also showed that \textsc{Teleoperation} resulted in a significantly higher (24.36\%) task load ($m=58.2$, $sd=15.483$) compared to \textsc{Monitoring} ($m=46.8$, $sd=18.715$). 
There were no interaction effects between \textsc{Interface} and \textsc{Scenario}.

\vspace{-0.03cm}
\paragraph{Simulator Sickness:} \textsc{VR} resulted in a significantly higher (43.45\%) simulator sickness ($m=41.6$, $sd=44.02$) compared to \textsc{Baseline} ($m=29$, $sd=36.77$). 
We did not find a significant effect of \textsc{Scenario} on simulator sickness, while comparing \textsc{Monitoring} ($m=39.27$, $sd=39.95$) to \textsc{Teleoperation} ($m=31.42$, $sd=41.76$). 
There were no interaction effects between \textsc{Interface} and \textsc{Scenario}. 

\vspace{-0.03cm}
\paragraph{Presence:} In \textsc{VR} interface, \textsc{Teleoperation} resulted in a significantly higher (15.97\%) ($p=0.01 $, $\eta^2p = 0.34$) presence ($m=3.34$, $sd=0.88$) compared to \textsc{Monitoring} ($m=2.88$, $sd=0.95$). 

\vspace{-0.03cm}
\paragraph{User Preferences:}

The responses from the preference questionnaire indicated that 87.5\% of the participants preferred \textsc{Baseline}, 56.25\% preferred \textsc{Monitoring}, and the rest 43.75\% preferred \textsc{Teleoperation}. 

Upon interviewing the participants, we got the following insights.
Six participants 
perceived a better performance and state of focus in VR, as they felt isolated in VR and were not distracted by the real surroundings. 
Three participants 
reported losing focus in \textsc{Baseline}, especially while monitoring. 
Eight participants 
mentioned having a higher eye strain due to the low resolution of VR.
While \textsc{Teleoperation}, they found it comparatively harder to detect the speed limit signs and obstacles in VR. 
Seven participants 
mentioned having more head movement in VR due to the small field of view of the HMD and peripheral blur. 
Eight participants 
mentioned that detecting the change in motion, i.e., the camera alerts felt the most difficult. 
Six participants 
felt more exhausted in \textsc{VR} due to the weight of the headset. 
Six participants 
reported a higher neck strain due to the position of \textsc{top}, the map screen. 
Five participants 
felt bored and underwhelmed while \textsc{Monitoring}, especially in \textsc{Baseline}.

In \textsc{Teleoperation}, one participant mentioned looking for alerts only during red traffic lights, which led to a perceived performance drop in monitoring tasks. 
Six participants 
reported feeling unsafe and insecure driving and monitoring the vehicles simultaneously. 
Four participants 
mentioned that a 3D perception of the driving environment, with stereoscopic cameras attached to the vehicles could lead to a better teleoperation experience. 
Four participants 
mentioned that the driving did not feel realistic due to the absence of other sensory feedback, like vibrations from the engine revving and sounds from the surroundings. 
Three participants 
mentioned that they could imagine using a VR headset for the monitoring and teleoperation tasks given improvements in resolution, peripheral sharpness, and a decrease in the weight of the headset. 

\vspace{-0.05cm}
\subsection{Discussion}
This study indicated, that the joint teleoperation and monitoring led to a dropped performance in terms of a 26.7\% higher miss rate and 24.36\% higher task load, hence it is not advisable to perform such joint tasks. 
It also indicated that the VR interface led to a dropped performance with 7.6\% more alerts missed in VR compared to the baseline. 
While driving, there was a 22.5\% higher deviation from the prescribed path in VR compared to the baseline. 
The task load in VR was 16.5\% higher and the simulator sickness was 43.45\% than the baseline. 
These results indicate the need to separate the monitoring and teleoperation tasks, the scope for further optimizations of the interface, and the need for a VR headset with better specifications.

\begin{figure}[t]
	\centering
	\includegraphics[width=0.95\linewidth]{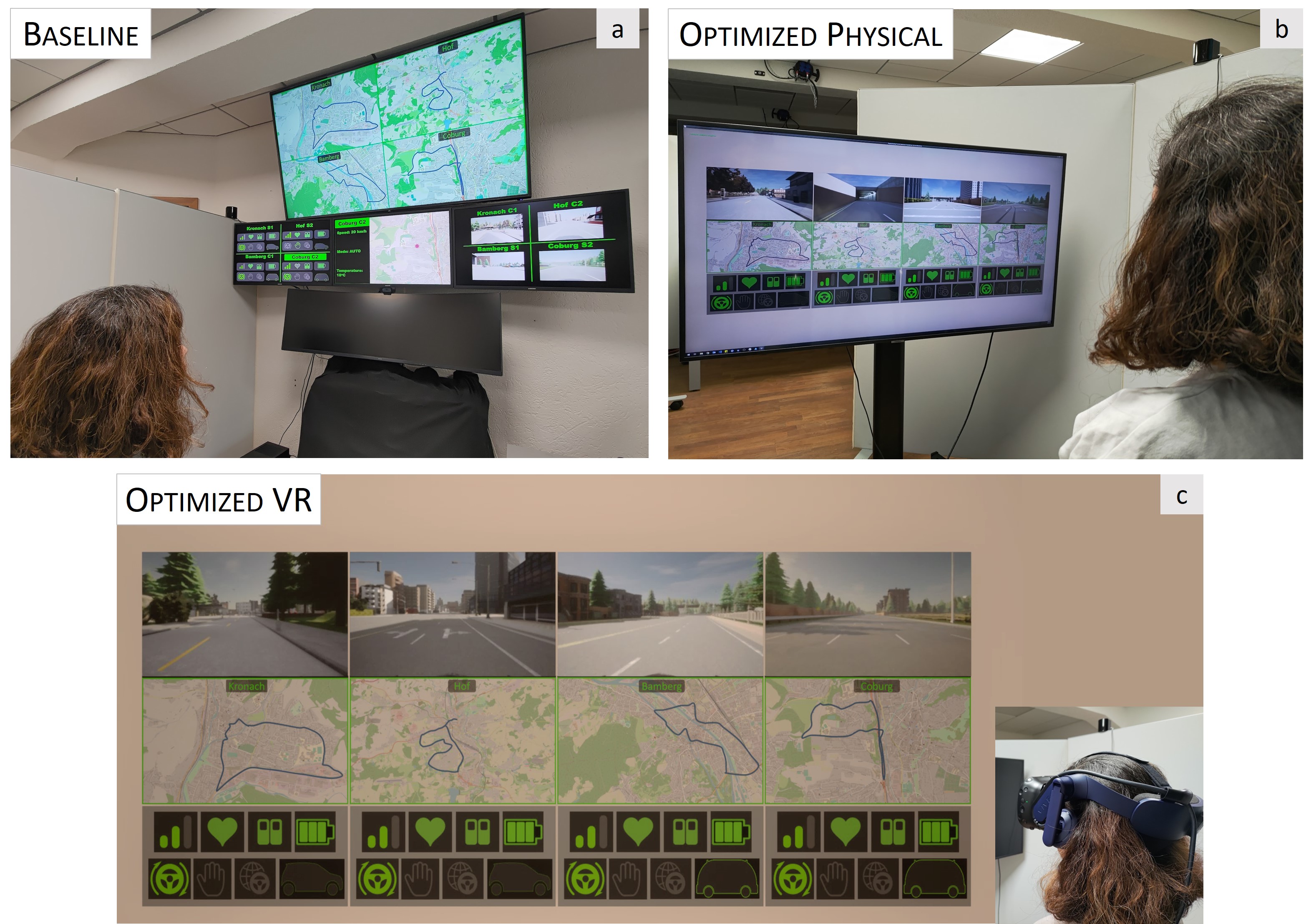}
        \vspace*{-0.2cm}
	\caption{Conditions of Study 2: a) The physical \textsc{Baseline} interface used in industry. b) The \textsc{Optimized Physical} interface. c) The \textsc{Optimized VR} interface. 
    }
	\label{fig:Study2-Bl-vs-Op}
         \vspace*{-0.6cm}

\end{figure}

\vspace{-0.1cm}
\section{Study 2: Evaluation of Optimized Layouts for Monitoring}
\label{sec: optimized-vr-study}

In Study 1, we quantified the negative effects of a joint teleoperation and monitoring task. These effects are so substantial, that we advise against performing both tasks simultaneously for a single operator (see also discussion section).
Still, for a single task (either monitoring or teleoperation) at a time or for switching between tasks consecutively, VR could provide potential benefits, such as providing task-dependent optimal screen layouts at no substantial monetary or infrastructure costs \cite{medeiros2022shielding}. 
Hence, in Study 2, we aimed at quantifying the effects of using a virtual screen setup optimized for monitoring only. To this end, we revised a screen layout for monitoring only, both for a 56-inch physical screen and a corresponding virtual screen, addressing the issues directly indicated by Study 1. The revised screen layout was chosen such that the out-of-view elements, peripheral blur, and fatigue-inducing screen distances causing extra head movements, were minimized.

Hence, the participants had to perform only the monitoring tasks (cf. \autoref{sec:monitoring-tasks}) in this user study. While having task-dependent monitor layouts is impractical using physical monitors, in VR, layout changes of virtual monitors can easily be achieved situation-dependent \cite{medeiros2022shielding}. Still, we included a task-optimized physical condition to be able to separate the effects of the screen layout and the presentation medium (monitor, VR HMD).

This user study was conducted as a 1$\times$3 within-subjects design. The independent variable was \textsc{Interface} with three levels, \textsc{Baseline}, \textsc{Optimized Physical}, and \textsc{Optimized VR}, resulting in three conditions. 
We measured the miss rate, reaction time for each alert, and error rate, as the objective data, and SUS, TLX, and SSQ, as the subjective data.
The order of the conditions was balanced. 
The procedure of this user study was the same as Study 1. 
Before each condition, we asked the participants to adjust the optimized layout to a height comfortable to their eyes. 
This study lasted for 90 minutes. 
We used a comparable procedure as that in \autoref{sec: participants-1} to get the minimum sample size. 
The participants had to be at least 18 years old to participate in the study. Twenty-four participants took part in the study (7 female, 17 male). The average age was 25.39 years ($sd=4.54$). Four participants had no experience with VR. Ten participants wore glasses during the user study. Nine participants among all participants had also participated in Study 1 (c.f. \autoref{sec:replication-study}). 

\vspace{-0.05cm}
\subsection{Apparatus}
The \textsc{Baseline} interface was the same as mentioned in section \autoref{section:apparatus-1}. We did not use the curved monitor for teleoperation, as we excluded the \textsc{Teleoperation} scenario in this study. The \textsc{Optimized Physical} interface was the optimized layout displayed on a Samsung 56-inch TV mounted on a height-adjustable stand. We used the same headset, HTC Vive Pro 2 for \textsc{Optimized VR} interface to keep the results of this user study comparable to Study 1. To seat the participants, we used a chair instead of the GTTRACK racing simulator cockpit, as the monitoring task did not demand such hardware. We maintained the same seat height from the ground as that with the GTTRACK racing simulator cockpit seat. The distance between the Samsung 56-inch TV and the participant was approximately 110 cm. We maintained the same in \textsc{Optimized VR}.

We changed the monitor layout to optimize the interface for a monitoring-only task, based on feedback from the participants of Study 1. 
This meant fitting the layout to a usable field of view, making individual information larger, e.g., the cameras, or smaller, e.g., the maps. The optimized layout is shown in \autoref{fig:Study2-Bl-vs-Op} (b) and (c). This layout was developed in UE 4.27.
We added a feature to adjust the vertical position of the optimized layout to a comfortable eye level, using the up and down arrow keys. 
To make this adjustment in \textsc{Optimized Physical}, the participants moved the optimized layout by a length of approximately 11 cm up or down within the TV screen with a wireless keyboard. 
For more adjustment, the TV was moved up or down physically on the stand.
In \textsc{Optimized VR}, the participants only used the up and down arrow keys for adjustment.

\vspace{-0.05cm}

\subsection{Results}
\begin{figure}[t]
	\centering
	\includegraphics[width=0.95\columnwidth]{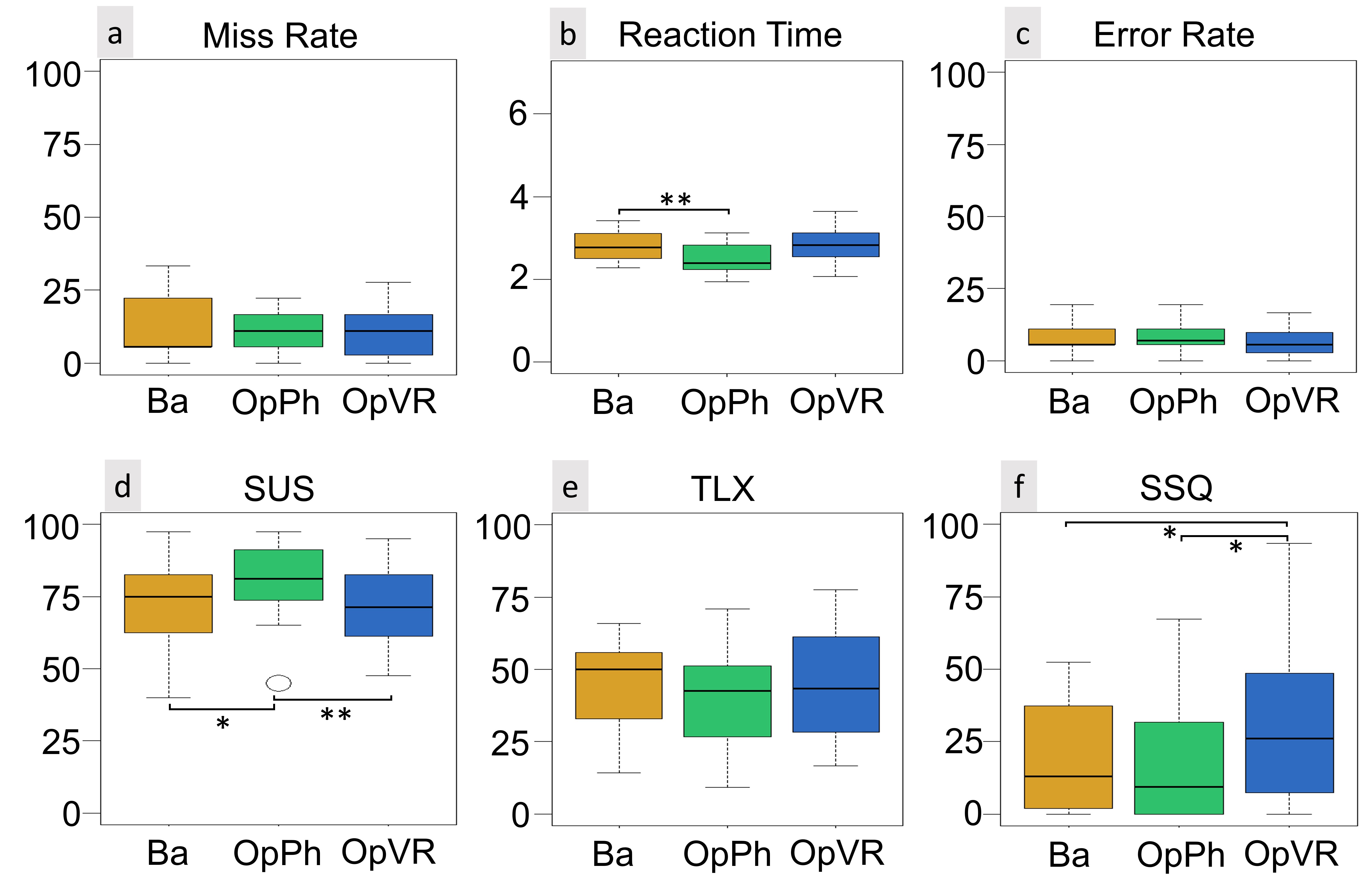}
	    \vspace{-0.4cm}
        \caption{Box plots for Study 2: a) Miss Rate. b) Reaction Time (in seconds). c) Error Rate. d) System Usability Scale. e) NASA Task Load Index. f) Simulator Sickness. Ba=\textsc{Baseline}, OpPh=\textsc{Optimized Physical}, OpVR=\textsc{Optimized VR}.
         The number of stars indicates the level of significance between the conditions (*** \textless 0.001 **  \textless 0.01 *  \textless 0.05).
         }
	\label{fig:optimized-vs-bl-plots}
 \vspace{-0.6cm}
\end{figure}

\begin{figure}[t]
	\centering
	\includegraphics[width=\linewidth]{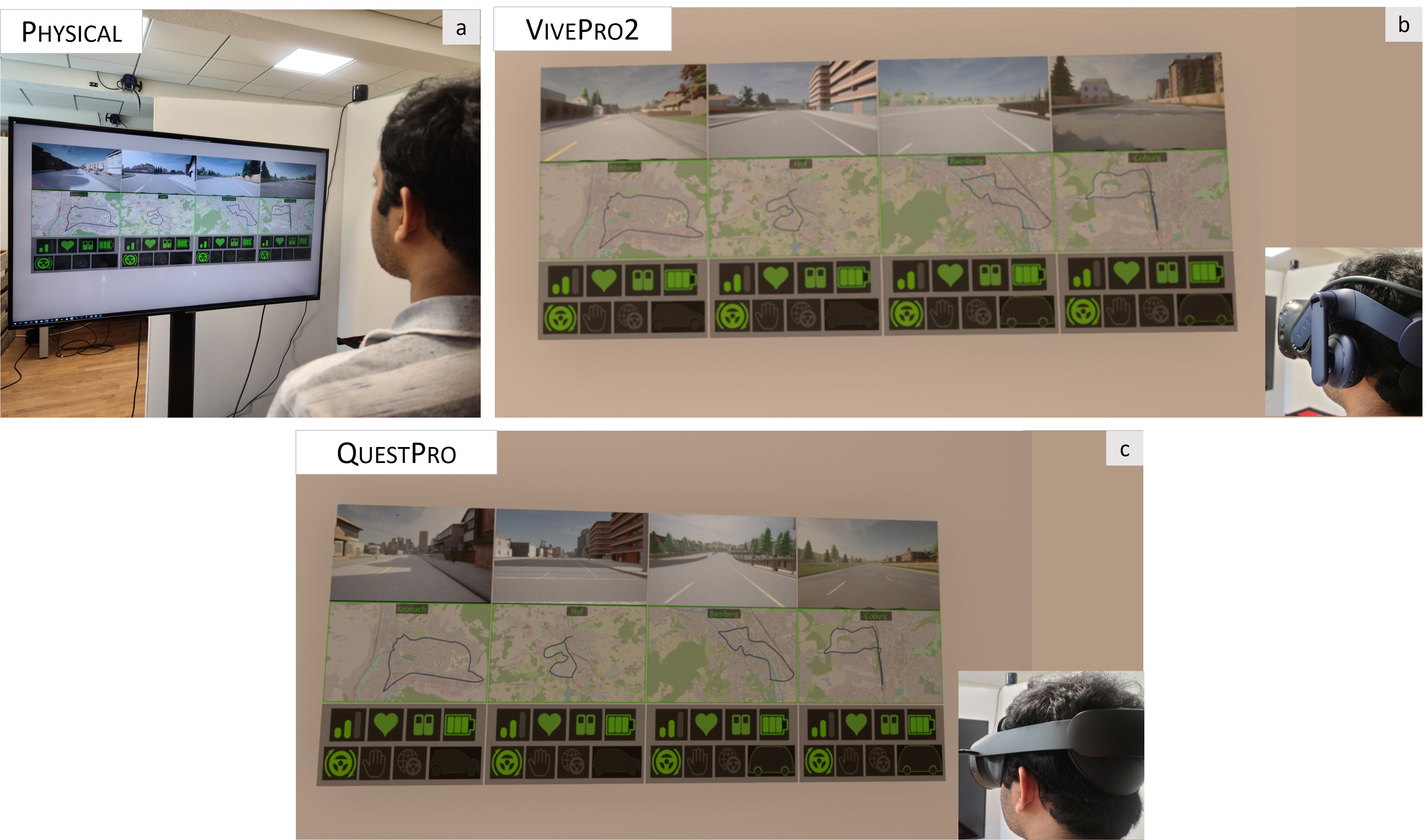}
        \vspace*{-0.6cm}
	\caption{Conditions of Study 3: a) The \textsc{Physical} interface. b) The \textsc{VivePro2} interface. c) The \textsc{QuestPro} interface. 
        }
        \vspace*{-0.2cm}
	\label{fig:Study3}
\end{figure}

We carried out similar inferential statistics as that in Study 1. 
The equivalence lower and upper bounds calculated using Cohen's $d_z$ with critical $t$ value for $n=20, \alpha=0.05$ for two-tails were $-0.47$ and $+0.47$ respectively.
Due to communication errors and confusion in one of the monitoring tasks, we had to remove the data of one participant. The task data of three other participants deviated by more than thrice the mean values, hence, we removed these outliers from all the metrics for our statistical analysis. One of these outliers was a participant from Study 1, but a worse performance from this participant ruled out any learning effects. 
The box plots for each dependent variable are shown in \autoref{fig:optimized-vs-bl-plots}. The results of the statistical tests are presented in \autoref{tab:resultsTableStudy2}. 

\begin{table}[t]
    \centering 
    \tiny
    \caption{RM-ANOVA results for Study 2. 
    d$f_1$ = d$f_{effect}$ and d$f_2$ = d$f_{error}$.
    }
    \setlength{\tabcolsep}{5pt}
        
        \begin{tabularx}{0.47\textwidth}{|C||c|c|c|c|c||c|c|c|c|c|}
            \hline
            &\multicolumn{5}{|c|}{Miss Rate} &\multicolumn{5}{|c|}{Reaction Time} 
            \\
            \cline{2-11}
            & d$f_1$ & d$f_2$ & F & p &  $\eta^2_p$
            & d$f_1$ & d$f_2$ & F & p &  $\eta^2_p$ 
            \\
            \hline
            \textsc{Interface} & $2$ & $38$ & $0.44$ & $0.647$ & $0.023$
                                  & $2$ & $38$ & $5.18$ & \cellcolor{lightgray}$0.01$ & $0.214$
                                 
                                 \\             
            \hline                    
        \end{tabularx}
        \\ 
        \begin{tabularx}{0.47\textwidth}{|C||c|c|c|c|c||c|c|c|c|c|}
            \hline
            &\multicolumn{5}{|c|}{Error Rate} &\multicolumn{5}{|c|}{System Usability}
            \\
            \cline{2-11}
            & d$f_1$ & d$f_2$ & F & p &  $\eta^2_p$
            & d$f_1$ & d$f_2$ & F & p &  $\eta^2_p$
            \\
            \hline
            \textsc{Interface} & $2$ & $38$& $0.286$ & $0.75$ & $0.015$ 
                                & $2$ & $38$ & $7.397$ & \cellcolor{lightgray}$<0.01$ & $0.28$
                                \\             
            \hline                    
        \end{tabularx}
        \\
        \begin{tabularx}{0.47\textwidth}{|C||c|c|c|c|c||c|c|c|c|c|}
            \hline
            &\multicolumn{5}{|c|}{Task Load} &\multicolumn{5}{|c|}{Simulator Sickness}\\
            \cline{2-11}
            & d$f_1$ & d$f_2$ & F & p &  $\eta^2_p$
            & d$f_1$ & d$f_2$ & F & p &  $\eta^2_p$
            \\
            \hline
            \textsc{Interface} & $2$ & $38$ & $1.881$ & $0.166$ & $0.09$ 
                                & $2$ & $38$ & $5.648$ & \cellcolor{lightgray}$<0.01$ & $0.23$
                                \\             
            \hline                    
        \end{tabularx}
        \label{tab:resultsTableStudy2}
        \vspace{-0.3cm}
\end{table}

\textsc{Optimized Physical} resulted in a significantly lower reaction time, 10.4\% lower than \textsc{Optimized VR} and 10.4\% lower than \textsc{Baseline}. It resulted in significantly higher system usability, 14.4\% higher than \textsc{Optimized VR} and 11.4\% higher than \textsc{Baseline}. \textsc{Optimized VR} resulted in significantly higher simulator sickness, 57.6\% higher than \textsc{Optimized Physical} and 60.5\% higher than \textsc{Baseline}. We did not find a significant effect of \textsc{Interface} for miss rate, error rate, and task load, and neither did we find an equivalence between either pair of \textsc{Interface} taken together.

\vspace{-0.05cm}
\paragraph{User Preferences:} 
The responses to the preference questionnaire indicated that 58.33\% of the participants preferred \textsc{Optimized Physical}, 25\% preferred \textsc{Baseline}, and 16.67\% preferred \textsc{Optimized VR}. 
In the interviews, fourteen participants 
mentioned preferring \textsc{Optimized VR} if the VR headset had better peripheral sharpness and was lighter. 
Three participants 
felt barely any difference between \textsc{Optimized VR} and \textsc{Optimized Physical}. 
Six participants liked \textsc{Optimized VR} but preferred \textsc{Optimized Physical} as there was no weight on their head, had lesser eye strain, the whole layout fit within the field of view, and was sharp enough. 
On the contrary, three other participants 
reported more eyestrain in \textsc{Optimized Physical} compared to \textsc{Optimized VR} and preferred the slight head movement with the VR HMD (due to the peripheral blur) instead. Six participants 
reported that the peripheral blur in \textsc{Optimized VR} led to overall more head movement than \textsc{Optimized Physical}, but not as much in \textsc{Baseline}. 
Five participants 
mentioned losing focus and concentration irrespective of the interface. 
Three participants 
reported being able to focus better in \textsc{Optimized VR} as they felt more isolated. 
Six participants 
mentioned difficulty in detecting camera alerts in the optimized layout due to the horizontal alignment of all camera views. 
Seven participants 
reported being able to focus better in \textsc{Baseline}, due to the entities categorized as maps, sensor data, and camera feed. 

Additionally, using a between-subjects variable separating the participants common to both studies and the new participants, we did not find an influence of prior participation on the results.

\vspace{-0.1cm}
\section{Study 3: Evaluation of Optimized Layouts with an Improved VR Headset}
\label{sec: follow-up-study}

Sixty-seven percent of the participants from Study 2 (cf. \autoref{sec: optimized-vr-study}), did not prefer the HTC Vive Pro 2 due to either its low resolution, peripheral blur, or weight. Hence, we conducted a follow-up user study to account for such limitations, by comparing the performance with a Meta Quest Pro, which is lighter and provides sharper views. We empirically determined that a Meta Quest Pro was better regarding these factors, which are important for our use case that demands longer use (less weight) and monitoring visually (sharp peripheries).

In this user study, the participants performed the monitoring task in the optimized layout again. This user study was conducted as a 1$\times$3 within-subjects design, with the independent variable \textsc{Interface} with three levels, \textsc{Physical}, \textsc{VivePro2}, \textsc{QuestPro}. The dependent variables, the procedure, and the requirement for participation were the same as that in Study 2.

Twenty-four participants took part in the study (7 female, 17 male). The average age was 24.4 years ($sd=3.6$). One participant had never used VR before. Ten participants wore glasses during the user study. Eight participants among all participants had participated in both Study 1 and Study 2, and 8 others among all participants had participated in Study 2 but not Study 1.

\vspace{-0.2cm}
\subsection{Results}

\begin{table}[t]
    \centering 
    \tiny
    \caption{RM-ANOVA results for Study 3. 
    d$f_1$ = d$f_{effect}$ and d$f_2$ = d$f_{error}$.
    }
    \setlength{\tabcolsep}{5pt}
        
        \begin{tabularx}{0.47\textwidth}{|C||c|c|c|c|c||c|c|c|c|c|}
            \hline
            &\multicolumn{5}{|c|}{Miss Rate} &\multicolumn{5}{|c|}{Reaction Time}
            \\
            \cline{2-11}
            & d$f_1$ & d$f_2$ & F & p &  $\eta^2_p$
            & d$f_1$ & d$f_2$ & F & p &  $\eta^2_p$ 
            \\
            \hline
            \textsc{Interface} & $2$ & $46$ & $1.25$ & $0.295$ & $0.052$
                                  & $2$ & $46$ & $1.77$ & $0.183$ & $0.071$
                                 \\             
            \hline                    
        \end{tabularx}
        \\ 
        \begin{tabularx}{0.47\textwidth}{|C||c|c|c|c|c||c|c|c|c|c|}
            \hline
            &\multicolumn{5}{|c|}{Error Rate} &\multicolumn{5}{|c|}{System Usability}
            \\
            \cline{2-11}
            & d$f_1$ & d$f_2$ & F & p &  $\eta^2_p$
            & d$f_1$ & d$f_2$ & F & p &  $\eta^2_p$
            \\
            \hline
            \textsc{Interface} & $1.5$ & $34.7$& $1.63$ & $0.21$ & $0.066$  
                                & $2$ & $44$ & $18.46$ & \cellcolor{lightgray}$<0.01$ & $0.46$
                                \\             
            \hline                    
        \end{tabularx}
        \\ 
        \begin{tabularx}{0.47\textwidth}{|C||c|c|c|c|c||c|c|c|c|c|}
            \hline
            &\multicolumn{5}{|c|}{Task Load} &\multicolumn{5}{|c|}{Simulator Sickness}
           \\
            \cline{2-11}
            & d$f_1$ & d$f_2$ & F & p &  $\eta^2_p$
            & d$f_1$ & d$f_2$ & F & p &  $\eta^2_p$
            \\
            \hline
            \textsc{Interface}  & $2$ & $46$ & $7.99$ & \cellcolor{lightgray}$0.001$ & $0.26$ 
                                & $2$ & $44$ & $16.79$ & \cellcolor{lightgray}$<0.01$ & $0.43$ 
                                \\             
            \hline                    
        \end{tabularx}
        \label{tab:resultsTableStudy3}
        \vspace{-0.5cm}
\end{table}

\begin{figure}[t]
	\centering 
	\includegraphics[width=\columnwidth]{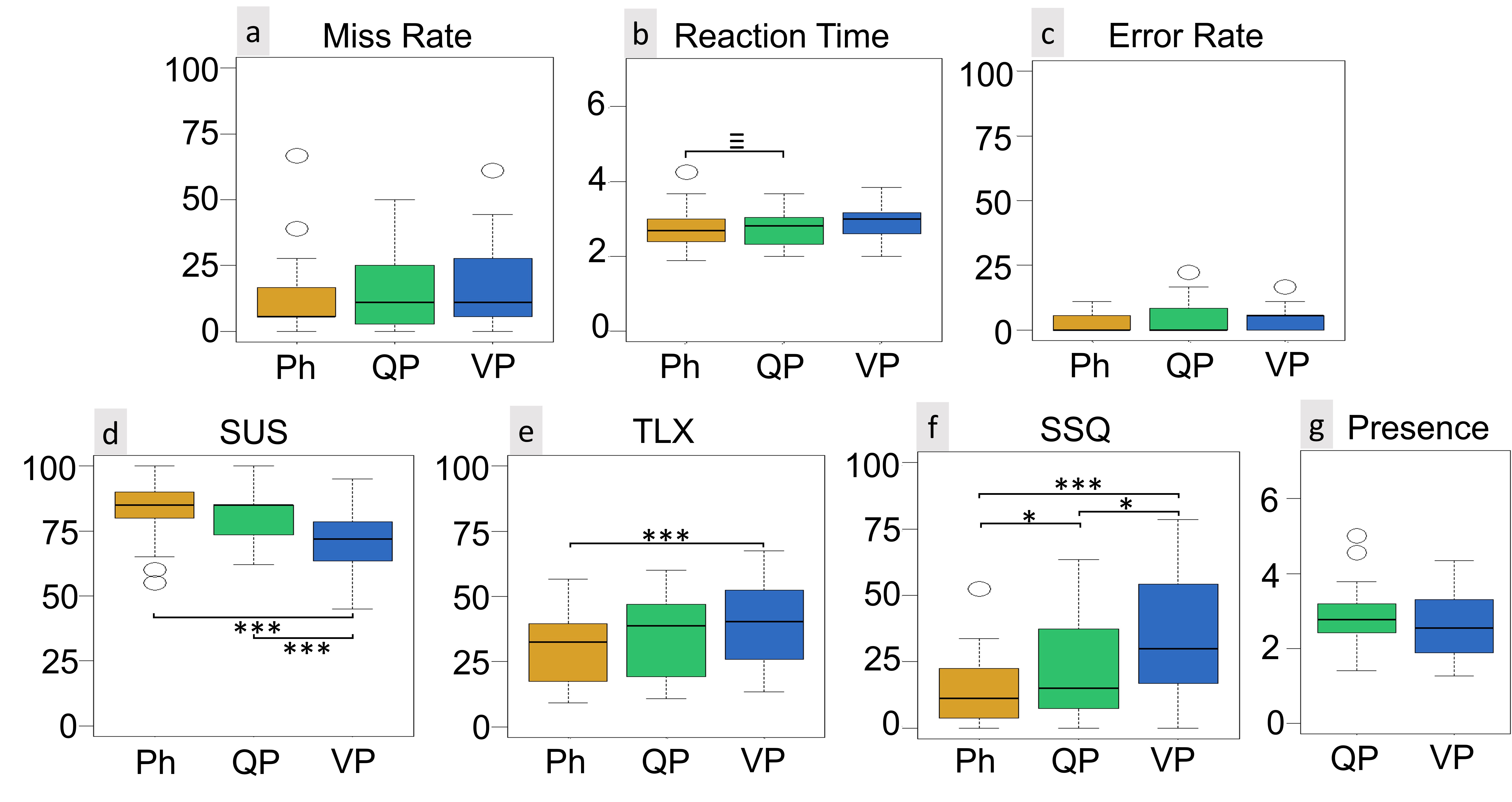}
	    \vspace{-0.6cm}
	\caption{Box plots for Study 3: a) Miss Rate. b) Reaction Time (in seconds). c) Error Rate. d) System Usability Scale. e) NASA Task Load Index. f) Simulator Sickness. g) Presence.
  The number of stars indicates the level of significance between the conditions (*** \textless 0.001 **  \textless 0.01 *  \textless 0.05). Equivalence between conditions is indicated by $\equiv$. }
	\label{fig:study3-analysis-plots}
	    \vspace{-0.6cm}
\end{figure}

We carried out inferential statistics as that in Study 1. 
The equivalence lower and upper bounds calculated using Cohen's $d_z$ with critical $t$ value for $n=24, \alpha=0.05$ for two-tails were $-0.42$ and $+0.42$ respectively.
The box plots for Study 3 are shown in \autoref{fig:study3-analysis-plots}. The results of the statistical tests are presented in \autoref{tab:resultsTableStudy3}. 

We did not find a significant effect but an equivalence effect of \textsc{Interface} on reaction time between \textsc{Physical} ($m=2.72$, $sd=0.57$) and \textsc{QuestPro} ($m=2.75$, $sd=0.47$), with the larger of the two p-values being $p=0.006$ and corresponding $t(23)=-2.729$. We did not find a significant effect of \textsc{Interface} on the miss rate or error rate. Neither did we find an equivalence effect between any two \textsc{Interface} taken together. 
\textsc{VivePro2} resulted in a significantly lower SUS score, 16.5\% lower than \textsc{Physical} and 12.4\% lower than \textsc{QuestPro}.
\textsc{Physical} resulted in a significantly lower task load, 32.5\% lower than \textsc{VivePro2} and 12.2\% lower than \textsc{QuestPro}.
\textsc{Physical} resulted in a significantly lower simulator sickness, 64.5\% lower than \textsc{VivePro2} and 42.9\% lower than \textsc{QuestPro}.
\textsc{QuestPro} resulted in a significantly lower simulator sickness (64.5\%) lower than \textsc{VivePro2}.
No significant difference in presence was found between \textsc{VivePro2} and \textsc{QuestPro}.

\paragraph{User Preferences:} The responses to the preference questionnaire indicated that 50\% of the participants preferred \textsc{QuestPro}, as it led to a perceived higher concentration and less distraction than \textsc{Physical} due to the isolation from reality. It felt lighter, less straining, and had better peripheral sharpness than \textsc{VivePro2} causing lesser head movements. 41.7\% preferred \textsc{Physical} and 8.3\% preferred \textsc{VivePro2}. 
Fifteen participants 
liked the peripheral view of \textsc{QuestPro} which caused lesser head movements and that it felt much lighter. 
Seven participants 
perceived more concentration through isolation with \textsc{QuestPro}. 
Six participants 
felt slightly uncomfortable with \textsc{QuestPro} due to the pressure on their heads and caused sweating. 
Three participants 
mentioned that they could not imagine using a VR headset for a long period. 
Sixteen participants perceived more head movement in \textsc{VivePro2} due to peripheral blur and felt more eye strain due to low resolution. 
Three participants 
felt more immersed in \textsc{VivePro2}, and mentioned a much better fit of this headset. 
Eight participants 
mentioned \textsc{Physical} felt easier due to all information lying within their field of view, and no weight or pressure on the head. 
Six participants 
mentioned drifting away from the task in \textsc{Physical}, as the real environment, at times took their attention away. 
Four participants 
did not feel much difference between \textsc{Physical} and \textsc{QuestPro}, but preferred \textsc{Physical} as it felt more comfortable due to no extra weight on their head.

In addition to the results of the 1$\times$3 within-subjects design, we investigated the effects of prior participation with a between-subjects variable \textsc{Common} separating the participants common to all studies, common to Study 2 and Study 3, and the new participants. 
We found that the participants common to all studies had significantly lower reaction times but did not find any interaction effect of \textsc{Interface} and \textsc{Common}. Hence they did not get better with a specific interface due to prior participation. 

\vspace{-0.1cm}
\section{Study Limitations}
\label{sec:limitations}
Our studies have limitations regarding the involved participants, the realism of the simulation, and the comparison of the technologies. 
First, the limited number of participants did not allow the investigation of smaller effects and subtle differences between the different conditions. 
Having a driver's license, all participants would be potentially qualified as operators for remote vehicles. Yet they were not trained for this job specifically and did not have any experience in this regard. 
The group of participants is not a representative sample of potential users, but a convenience sample. 
Additionally, we did not find that the repetition of participants in Study 2 and Study 3 did not influence the performance. However, the statistical power of this test was low ($1-\beta = 0.133$ for Study 2 and $1-\beta = 0.198$ for Study 3). Future work should substantially increase the sample size when including participants across multiple studies.

Second, we tried to model a real operation environment as closely as possible. Still, we had to compromise to keep the study feasible. For safety reasons, we could not operate real vehicles on a street but had to work with a simplified simulation. This takes away pressure not to make any mistakes but also reduces the sources of mistakes and task difficulty (e.g., no pedestrians or other traffic on the road). 
In our environment, the latency for teleoperation was minimal. However, in real-world teleoperation latency can have a negative impact on operating a remote car. The monitoring tasks were also simplified (e.g., triggering one alert at a time).

Third, a comparison between different technologies and representations is hard to design in a completely fair way. 
For example, the novelty effect of the interface could lead to higher motivation but also make the completion of the tasks harder. 
Furthermore, some effects might be just caused by the current technology limitations. 
For instance, as remarked by participants and highlighted by prior work \cite{pavanatto2021we, biener2022quantifying}, the resolution of most consumer-oriented VR HMDs cannot yet compete with the ones of physical monitors.
Wearing the HMD, the perception of alerts in the periphery could have been affected by the restricted field of view, or visual artifacts in the periphery. 
However, our analysis of the peripheral and central alerts indicates no large difference in the miss rates between the two. Future work should carefully study the impact of peripheral sharpness on task performance. 
Further, the weight and particular ergonomics of HMDs might impact the study results. In our studies, we only could quantify the combined effects of each display device. Future work could consider the influence of various factors individually.

\vspace{-0.2cm}
\section{Discussion}
Finally, we reflect on the results of the three studies jointly. We discuss implications for practical application and future research in the context of remote vehicle operation. 
As technology regarding autonomous driving and VR is evolving rapidly, we also try to capture technical solutions available in the near future.

\vspace{-0.05cm}
\paragraph*{Joint Monitoring and Teleoperation Mode.} 
Joint monitoring and teleoperation would be more cost-efficient and could lead to earlier adoption of level-4 autonomous driving. Such a dual operation is supported by the current industrial systems. With these ideas in mind, Study 1 integrated both monitoring and teleoperation tasks in two of the four conditions.
However, the additional teleoperation task led to higher miss rates of alerts and perceived insecurity while driving. 
While this result was expected, to the best of our knowledge, our study is the first that quantifies this difference (in our setting more than 25\% missed alerts for each interface). Hence, we strongly recommend not assigning joint monitoring and teleoperation to a single operator. 
If multiple operators are available per shift, flexible models for switching between monitoring and teleoperation roles might be feasible. This could also provide some variation in a potentially monotonous monitoring job. 
For such flexible switching between roles, a VR setup would allow more flexibility or both modes can complement each other. For instance, a physical interface could be used for monitoring and a VR interface for teleoperation.

\vspace{-0.05cm}
\paragraph*{Physical and VR Interfaces in Comparison.} 
The results comparing physical and VR interfaces indicate that the tested VR interfaces are secondary options: The physical interfaces outperform them regarding miss rate, deviation from the path (Study 1), task load, and simulator sickness (Study 1, Study 3) and reaction time, system usability, task load, and simulator sickness (Study 2). 
If VR should still be used (for instance, because of space constraints and mobility of the setup), we recommend making the operator's tasks as simple as possible for both monitoring and teleoperation. For instance, using particularly clear visual encodings for monitoring and assistance systems for driving. 
We also recommend using newer-generation HMDs with better specifications and ergonomics. They seem to leverage a benefit like better concentration through isolation, among others. 
Moreover, work sessions in VR should be shorter (task load, simulator sickness). 
These results align with those by Biener et al. \cite{biener2022quantifying}, who quantified the effects of working prolonged time with VR HMDs compared to work using physical monitors. 

\vspace{-0.06cm}
\paragraph*{VR Future Perspectives.} 
Current limitations of the technology and constraints of the experiment design have limited the used potential, specifically of the VR interfaces in our studies (cf. \autoref{sec:limitations}). 
With enlarged fields of view, better resolution, lesser weight, peripheral view, and larger \textit{sweet spots} for sharp vision, such biases could partly be reduced merely through technological progress.
This can already be seen in the user preferences of Study 3 with Meta Quest Pro.
For example, we confirm a higher perceived resolution of the Meta Quest Pro, potentially due to its pancake lenses subjectively. Still, a formal comparison of center vs. peripheral resolutions of Fresnel vs. pancake lenses in HMDs like the HTC Vive Pro 2 and Meta Quest Pro should be explored in the future.
Hence, replicating the study with next-generation HMDs is advisable (c.f., Schneider et al. \cite{schneider2021accuracy} for the case of finger tracking accuracy.
Moreover, to provide comparable designs and limit differences between study conditions in Study 1, the VR interface replicates the physical setup as best as possible without optimizing it for the VR environment. 
While some optimization is already applied in Study 2 and Study 3, further optimizations should be considered in future studies as proposed in prior work on supporting knowledge workers in VR \cite{biener2020breaking}. 
This could leverage more VR specifics at the cost of comparability, for instance arranging and sizing screens even more flexibly and dynamically, enlarging the visual encodings if they are depicted in the periphery of the HMD, to improve legibility.
Context-specific tasks instead of placeholder tasks should also be tested.
Further, the replication of current vehicle hardware limited us to using 2D monoscopic (simulated) cameras. 
If 360$^{\circ}$ images \cite{gafert2022teleoperationstation} and stereo images \cite{shen2016teleoperation} are available, the VR interface might provide a more immersive scene representation. This could also improve the situation awareness in teleoperation tasks. 
Regarding monitoring scenarios, especially spatial views using 3D maps and city models could provide better geographic context and overview in VR, similar as tested for autonomous vessels in a search and rescue scenario by Lager and Topp~\cite{10.1016/j.ifacol.2019.08.104}.

\vspace{-0.15cm}
\section{Conclusion}
We contributed three user studies, quantifying the performance of and subjective feedback for a VR-based system with an existing monitoring and teleoperation system, which is in industrial use today. 
Our studies indicate that when simply replicating physical interfaces in VR, the costs (in terms of perceptual and ergonomic issues) induced by the VR system outweigh potential benefits (such as better concentration through isolation). 
Further, we quantified the performance degradation in joint teleoperation and monitoring tasks independent of the interface (which were higher than 25\% in terms of missed alerts). 
Hence, we explicitly advise against promoting such joint tasks in the future. Finally, we are confident, that potential software-related (such as stereoscopic vision, flexible, situation-dependent screen layouts) and hardware-related (such as lighter headsets with a larger usable field of view) might outweigh the current costs of using VR for monitoring and teleoperation of autonomous vehicles in the future.

\acknowledgments{
This work was supported by the 5G-5GKC project, funded by the Federal Ministry for Transport and Digital Infrastructure, Germany (funding number: 165GU102C). The authors wish to thank Valeo for supporting the replication of their current physical setup, the participants of the experiment, and the anonymous reviewers for their feedback.
}

\bibliographystyle{abbrv-doi}

\balance
\bibliography{template}

\begin{thebibliography}{10}

\bibitem{bednarz2011applications}
T.~Bednarz, C.~James, C.~Caris, K.~Haustein, M.~Adcock, and C.~Gunn.
\newblock Applications of networked virtual reality for tele-operation and
  tele-assistance systems in the mining industry.
\newblock In {\em Proceedings of the 10th International Conference on Virtual
  Reality Continuum and Its Applications in Industry}, pp. 459--462, 2011. doi:
  {{%
10\hspace{.1pt}\discretionary{.}{%
}{.}\hspace{.4pt}1145\discretionary{/}{%
}{/}2087756\hspace{.1pt}\discretionary{.}{%
}{.}\hspace{.4pt}2087845}}


\bibitem{bergroth2018use}
J.~D. Bergroth, H.~M. Koskinen, and J.~O. Laarni.
\newblock Use of immersive {3-D} virtual reality environments in control room
  validations.
\newblock {\em Nuclear Technology}, 202(2-3):278--289, 2018. doi: {{%
10\hspace{.1pt}\discretionary{.}{%
}{.}\hspace{.4pt}1080\discretionary{/}{%
}{/}00295450\hspace{.1pt}\discretionary{.}{%
}{.}\hspace{.4pt}2017\hspace{.1pt}\discretionary{.}{%
}{.}\hspace{.4pt}1420335}}


\bibitem{biener2022quantifying}
V.~Biener, S.~Kalamkar, N.~Nouri, E.~Ofek, M.~Pahud, J.~J. Dudley, J.~Hu, P.~O.
  Kristensson, M.~Weerasinghe, K.~{\v{C}}. Pucihar, et~al.
\newblock Quantifying the effects of working in vr for one week.
\newblock {\em IEEE Transactions on Visualization and Computer Graphics},
  28(11):3810--3820, 2022.

\bibitem{biener2020breaking}
V.~Biener, D.~Schneider, T.~Gesslein, A.~Otte, B.~Kuth, P.~O. Kristensson,
  E.~Ofek, M.~Pahud, and J.~Grubert.
\newblock Breaking the screen: Interaction across touchscreen boundaries in
  virtual reality for mobile knowledge workers.
\newblock {\em arXiv preprint arXiv:2008.04559}, 2020.

\bibitem{blanca2023non}
M.~J. Blanca~Mena, J.~Arnau~Gras, F.~J. Garc{\'\i}a~de Castro,
  R.~Alarc{\'o}n~Postigo, R.~Bono~Cabr{\'e}, et~al.
\newblock Non-normal data in repeated measures anova: impact on type i error
  and power.
\newblock {\em Psicothema}, 2023.

\bibitem{brooke1996sus}
J.~Brooke et~al.
\newblock {SUS}---a quick and dirty usability scale.
\newblock {\em Usability Evaluation in Industry}, 189(194):4--7, 1996. doi: {{%
10\hspace{.1pt}\discretionary{.}{%
}{.}\hspace{.4pt}1201\discretionary{/}{%
}{/}9781498710411\discretionary{%
}{-}{-}35}}


\bibitem{colley2022effects}
M.~Colley, E.~Bajrovic, and E.~Rukzio.
\newblock Effects of pedestrian behavior, time pressure, and repeated exposure
  on crossing decisions in front of automated vehicles equipped with external
  communication.
\newblock In {\em Proceedings of the 2022 CHI Conference on Human Factors in
  Computing Systems}, pp. 1--11, 2022.

\bibitem{colley2020effect}
M.~Colley, C.~Br{\"a}uner, M.~Lanzer, M.~Walch, M.~Baumann, and E.~Rukzio.
\newblock Effect of visualization of pedestrian intention recognition on trust
  and cognitive load.
\newblock In {\em 12th International Conference on Automotive User Interfaces
  and Interactive Vehicular Applications}, pp. 181--191, 2020.

\bibitem{corbin1990grounded}
J.~M. Corbin and A.~Strauss.
\newblock Grounded theory research: Procedures, canons, and evaluative
  criteria.
\newblock {\em Qualitative sociology}, 13(1):3--21, 1990.

\bibitem{Dosovitskiy17}
A.~Dosovitskiy, G.~Ros, F.~Codevilla, A.~Lopez, and V.~Koltun.
\newblock {CARLA}: {An} open urban driving simulator.
\newblock In {\em Proceedings of the 1st Annual Conference on Robot Learning},
  pp. 1--16, 2017.

\bibitem{ens2014personal}
B.~M. Ens, R.~Finnegan, and P.~P. Irani.
\newblock The personal cockpit: a spatial interface for effective task
  switching on head-worn displays.
\newblock In {\em Proceedings of the SIGCHI Conference on Human Factors in
  Computing Systems}, pp. 3171--3180, 2014.

\bibitem{fabris2021immersive}
E.~J. Fabris, V.~A. Sangalli, L.~P. Soares, and M.~S. Pinho.
\newblock Immersive telepresence on the operation of unmanned vehicles.
\newblock {\em International Journal of Advanced Robotic Systems},
  18(1):1729881420978544, 2021.

\bibitem{faul2007g}
F.~Faul, E.~Erdfelder, A.-G. Lang, and A.~Buchner.
\newblock G* power 3: A flexible statistical power analysis program for the
  social, behavioral, and biomedical sciences.
\newblock {\em Behavior research methods}, 39(2):175--191, 2007.

\bibitem{gafert2022teleoperationstation}
M.~Gafert, A.~G. Mirnig, P.~Fr{\"o}hlich, and M.~Tscheligi.
\newblock {TeleOperationStation}: {XR}-exploration of user interfaces for
  remote automated vehicle operation.
\newblock In {\em CHI Conference on Human Factors in Computing Systems Extended
  Abstracts}, pp. 1--4, 2022. doi: {{%
10\hspace{.1pt}\discretionary{.}{%
}{.}\hspace{.4pt}1145\discretionary{/}{%
}{/}3491101\hspace{.1pt}\discretionary{.}{%
}{.}\hspace{.4pt}3519882}}


\bibitem{georg2019adaptable}
J.-M. Georg and F.~Diermeyer.
\newblock An adaptable and immersive real time interface for resolving system
  limitations of automated vehicles with teleoperation.
\newblock In {\em 2019 IEEE International Conference on Systems, Man and
  Cybernetics (SMC)}, pp. 2659--2664. IEEE, 2019. doi: {{%
10\hspace{.1pt}\discretionary{.}{%
}{.}\hspace{.4pt}1109\discretionary{/}{%
}{/}smc\hspace{.1pt}\discretionary{.}{%
}{.}\hspace{.4pt}2019\hspace{.1pt}\discretionary{.}{%
}{.}\hspace{.4pt}8914306}}


\bibitem{10.1109/itsc.2018.8569408}
J.-M. Georg, J.~Feiler, F.~Diermeyer, and M.~Lienkamp.
\newblock Teleoperated driving, a key technology for automated driving?
  {C}omparison of actual test drives with a head mounted display and
  conventional monitors.
\newblock In {\em 2018 21st International Conference on Intelligent
  Transportation Systems (ITSC)}, 2018. doi: {{%
10\hspace{.1pt}\discretionary{.}{%
}{.}\hspace{.4pt}1109\discretionary{/}{%
}{/}itsc\hspace{.1pt}\discretionary{.}{%
}{.}\hspace{.4pt}2018\hspace{.1pt}\discretionary{.}{%
}{.}\hspace{.4pt}8569408}}


\bibitem{graf2020user}
G.~Graf and H.~Hussmann.
\newblock User requirements for remote teleoperation-based interfaces.
\newblock In {\em 12th International Conference on Automotive User Interfaces
  and Interactive Vehicular Applications}, pp. 85--88, 2020. doi: {{%
10\hspace{.1pt}\discretionary{.}{%
}{.}\hspace{.4pt}1145\discretionary{/}{%
}{/}3409251\hspace{.1pt}\discretionary{.}{%
}{.}\hspace{.4pt}3411730}}


\bibitem{graf2020design}
G.~Graf, H.~Palleis, and H.~Hussmann.
\newblock A design space for advanced visual interfaces for teleoperated
  autonomous vehicles.
\newblock In {\em Proceedings of the International Conference on Advanced
  Visual Interfaces}, pp. 1--3, 2020. doi: {{%
10\hspace{.1pt}\discretionary{.}{%
}{.}\hspace{.4pt}1145\discretionary{/}{%
}{/}3399715\hspace{.1pt}\discretionary{.}{%
}{.}\hspace{.4pt}3399942}}


\bibitem{grubert2018office}
J.~Grubert, E.~Ofek, M.~Pahud, and P.~O. Kristensson.
\newblock The office of the future: Virtual, portable, and global.
\newblock {\em IEEE computer graphics and applications}, 38(6):125--133, 2018.

\bibitem{hart1988development}
S.~G. Hart and L.~E. Staveland.
\newblock Development of {NASA-TLX (Task Load Index)}: Results of empirical and
  theoretical research.
\newblock In {\em Advances in psychology}, vol.~52, pp. 139--183. Elsevier,
  1988. doi: {{%
10\hspace{.1pt}\discretionary{.}{%
}{.}\hspace{.4pt}1016\discretionary{/}{%
}{/}S0166\discretionary{%
}{-}{-}4115\discretionary{%
}{(}{(}08\discretionary{)}{%
}{)}62386\discretionary{%
}{-}{-}9}}


\bibitem{hensch2020effects}
A.-C. Hensch, N.~Rauh, C.~Schmidt, S.~Hergeth, F.~Naujoks, J.~F. Krems, and
  A.~Keinath.
\newblock Effects of secondary tasks and display position on glance behavior
  during partially automated driving.
\newblock {\em Transportation research part F: traffic psychology and
  behaviour}, 68:23--32, 2020.

\bibitem{hosseini2016enhancing}
A.~Hosseini and M.~Lienkamp.
\newblock Enhancing telepresence during the teleoperation of road vehicles
  using {HMD}-based mixed reality.
\newblock In {\em 2016 IEEE Intelligent Vehicles Symposium (IV)}, pp.
  1366--1373. IEEE, 2016. doi: {{%
10\hspace{.1pt}\discretionary{.}{%
}{.}\hspace{.4pt}1109\discretionary{/}{%
}{/}ivs\hspace{.1pt}\discretionary{.}{%
}{.}\hspace{.4pt}2016\hspace{.1pt}\discretionary{.}{%
}{.}\hspace{.4pt}7535568}}


\bibitem{sae2018taxonomy}
S.~International.
\newblock Taxonomy and definitions for terms related to driving automation
  systems for on-road motor vehicles.
\newblock {\em SAE international}, 4970(724):1--5, 2018.

\bibitem{jansen2023autovis}
P.~Jansen, J.~Britten, A.~H{\"a}usele, T.~Segschneider, M.~Colley, and
  E.~Rukzio.
\newblock Autovis: Enabling mixed-immersive analysis of automotive user
  interface interaction studies.
\newblock {\em arXiv preprint arXiv:2302.10531}, 2023.

\bibitem{ji2020design}
Y.~Ji.
\newblock Design of ship visual communication teleoperation system based on
  virtual reality.
\newblock {\em Journal of Coastal Research}, 103(SI):975--978, 2020.

\bibitem{kalinov2021warevr}
I.~Kalinov, D.~Trinitatova, and D.~Tsetserukou.
\newblock {WareVR}: Virtual reality interface for supervision of autonomous
  robotic system aimed at warehouse stocktaking.
\newblock In {\em 2021 IEEE International Conference on Systems, Man, and
  Cybernetics (SMC)}, pp. 2139--2145. IEEE, 2021. doi: {{%
10\hspace{.1pt}\discretionary{.}{%
}{.}\hspace{.4pt}1109\discretionary{/}{%
}{/}smc52423\hspace{.1pt}\discretionary{.}{%
}{.}\hspace{.4pt}2021\hspace{.1pt}\discretionary{.}{%
}{.}\hspace{.4pt}9659133}}


\bibitem{kennedy1993simulator}
R.~S. Kennedy, N.~E. Lane, K.~S. Berbaum, and M.~G. Lilienthal.
\newblock Simulator sickness questionnaire: An enhanced method for quantifying
  simulator sickness.
\newblock {\em The International Journal of Aviation Psychology},
  3(3):203--220, 1993. doi: {{%
10\hspace{.1pt}\discretionary{.}{%
}{.}\hspace{.4pt}1207\discretionary{/}{%
}{/}s15327108ijap0303\_3}}


\bibitem{kettwich2021teleoperation}
C.~Kettwich, A.~Schrank, and M.~Oehl.
\newblock Teleoperation of highly automated vehicles in public transport:
  User-centered design of a human-machine interface for remote-operation and
  its expert usability evaluation.
\newblock {\em Multimodal Technologies and Interaction}, 5(5):26, 2021. doi:
  {{%
10\hspace{.1pt}\discretionary{.}{%
}{.}\hspace{.4pt}3390\discretionary{/}{%
}{/}mti5050026}}


\bibitem{10.1016/j.ifacol.2019.08.104}
M.~Lager and E.~A. Topp.
\newblock Remote supervision of an autonomous surface vehicle using virtual
  reality.
\newblock {\em IFAC-PapersOnLine}, 52(8):387--392, 2019. doi: {{%
10\hspace{.1pt}\discretionary{.}{%
}{.}\hspace{.4pt}1016\discretionary{/}{%
}{/}j\hspace{.1pt}\discretionary{.}{%
}{.}\hspace{.4pt}ifacol\hspace{.1pt}\discretionary{.}{%
}{.}\hspace{.4pt}2019\hspace{.1pt}\discretionary{.}{%
}{.}\hspace{.4pt}08\hspace{.1pt}\discretionary{.}{%
}{.}\hspace{.4pt}104}}


\bibitem{lakens2018equivalence}
D.~Lakens, A.~M. Scheel, and P.~M. Isager.
\newblock Equivalence testing for psychological research: A tutorial.
\newblock {\em Advances in Methods and Practices in Psychological Science},
  1(2):259--269, 2018.

\bibitem{li2021rear}
J.~Li, C.~George, A.~Ngao, K.~Holl{\"a}nder, S.~Mayer, and A.~Butz.
\newblock Rear-seat productivity in virtual reality: Investigating vr
  interaction in the confined space of a car.
\newblock {\em Multimodal Technologies and Interaction}, 5(4):15, 2021.

\bibitem{lischke2016screen}
L.~Lischke, S.~Mayer, K.~Wolf, N.~Henze, H.~Reiterer, and A.~Schmidt.
\newblock Screen arrangements and interaction areas for large display work
  places.
\newblock In {\em Proceedings of the 5th ACM International Symposium on
  Pervasive Displays}, pp. 228--234, 2016.

\bibitem{mcgill2020expanding}
M.~Mcgill, A.~Kehoe, E.~Freeman, and S.~Brewster.
\newblock Expanding the bounds of seated virtual workspaces.
\newblock {\em ACM Transactions on Computer-Human Interaction (TOCHI)},
  27(3):1--40, 2020.

\bibitem{mcgill2020challenges}
M.~McGill, J.~Williamson, A.~Ng, F.~Pollick, and S.~Brewster.
\newblock Challenges in passenger use of mixed reality headsets in cars and
  other transportation.
\newblock {\em Virtual Reality}, 24:583--603, 2020.

\bibitem{medeiros2022shielding}
D.~Medeiros, M.~McGill, A.~Ng, R.~McDermid, N.~Pantidi, J.~Williamson, and
  S.~Brewster.
\newblock From shielding to avoidance: Passenger augmented reality and the
  layout of virtual displays for productivity in shared transit.
\newblock {\em IEEE Transactions on Visualization and Computer Graphics},
  28(11):3640--3650, 2022.

\bibitem{morra2019building}
L.~Morra, F.~Lamberti, F.~G. Prattic{\'o}, S.~La~Rosa, and P.~Montuschi.
\newblock Building trust in autonomous vehicles: Role of virtual reality
  driving simulators in hmi design.
\newblock {\em IEEE Transactions on Vehicular Technology}, 68(10):9438--9450,
  2019.

\bibitem{neumeier2018way}
S.~Neumeier, N.~Gay, C.~Dannheim, and C.~Facchi.
\newblock On the way to autonomous vehicles teleoperated driving.
\newblock In {\em AmE 2018-Automotive meets Electronics; 9th GMM-Symposium},
  pp. 1--6. VDE, 2018.

\bibitem{paredes2018driving}
P.~E. Paredes, S.~Balters, K.~Qian, E.~L. Murnane, F.~Ord{\'o}{\~n}ez, W.~Ju,
  and J.~A. Landay.
\newblock Driving with the fishes: Towards calming and mindful virtual reality
  experiences for the car.
\newblock {\em Proceedings of the ACM on Interactive, Mobile, Wearable and
  Ubiquitous Technologies}, 2(4):1--21, 2018.

\bibitem{pavanatto2021we}
L.~Pavanatto, C.~North, D.~A. Bowman, C.~Badea, and R.~Stoakley.
\newblock Do we still need physical monitors? an evaluation of the usability of
  ar virtual monitors for productivity work.
\newblock In {\em 2021 IEEE Virtual Reality and 3D User Interfaces (VR)}, pp.
  759--767. IEEE, 2021.

\bibitem{riegler2021systematic}
A.~Riegler, A.~Riener, and C.~Holzmann.
\newblock A systematic review of virtual reality applications for automated
  driving: 2009--2020.
\newblock {\em Frontiers in human dynamics}, 3:689856, 2021.

\bibitem{schneider2021accuracy}
D.~Schneider, V.~Biener, A.~Otte, T.~Gesslein, P.~Gagel, C.~Campos,
  K.~{\v{C}}opi{\v{c}}~Pucihar, M.~Kljun, E.~Ofek, M.~Pahud, et~al.
\newblock Accuracy evaluation of touch tasks in commodity virtual and augmented
  reality head-mounted displays.
\newblock In {\em Proceedings of the 2021 ACM Symposium on Spatial User
  Interaction}, pp. 1--11, 2021.

\bibitem{schubert2003sense}
T.~W. Schubert.
\newblock The sense of presence in virtual environments: A three-component
  scale measuring spatial presence, involvement, and realness.
\newblock {\em Z. f{\"u}r Medienpsychologie}, 15(2):69--71, 2003.

\bibitem{shen2016teleoperation}
X.~Shen, Z.~J. Chong, S.~Pendleton, G.~M. James~Fu, B.~Qin, E.~Frazzoli, and
  M.~H. Ang.
\newblock Teleoperation of on-road vehicles via immersive telepresence using
  off-the-shelf components.
\newblock In {\em Intelligent Autonomous Systems 13: Proceedings of the 13th
  International Conference IAS-13}, pp. 1419--1433. Springer, 2016. doi: {{%
10\hspace{.1pt}\discretionary{.}{%
}{.}\hspace{.4pt}1007\discretionary{/}{%
}{/}978\discretionary{%
}{-}{-}3\discretionary{%
}{-}{-}319\discretionary{%
}{-}{-}08338\discretionary{%
}{-}{-}4\_102}}


\bibitem{tsigkounis2021monitoring}
K.~Tsigkounis, A.~Komninos, N.~Politis, and J.~Garofalakis.
\newblock Monitoring maritime industry 4.0 systems through {VR} environments.
\newblock In {\em CHI Greece 2021: 1st International Conference of the ACM
  Greek SIGCHI Chapter}, pp. 1--8, 2021. doi: {{%
10\hspace{.1pt}\discretionary{.}{%
}{.}\hspace{.4pt}1145\discretionary{/}{%
}{/}3489410\hspace{.1pt}\discretionary{.}{%
}{.}\hspace{.4pt}3489429}}


\bibitem{whitney2018ros}
D.~Whitney, E.~Rosen, D.~Ullman, E.~Phillips, and S.~Tellex.
\newblock Ros reality: A virtual reality framework using consumer-grade
  hardware for ros-enabled robots.
\newblock In {\em 2018 IEEE/RSJ International Conference on Intelligent Robots
  and Systems (IROS)}, pp. 1--9. IEEE, 2018.

\bibitem{ye2020risks}
Y.~Ye, S.~Wong, Y.~Li, and Y.~Lau.
\newblock Risks to pedestrians in traffic systems with unfamiliar driving
  rules: A virtual reality approach.
\newblock {\em Accident Analysis \& Prevention}, 142:105565, 2020.

\end{thebibliography}
\end{document}